\documentstyle[12pt,a4,epsf]{article}

\title{Primordial black holes as a source of extremely high energy cosmic rays}
\author{Aur\'elien Barrau}
\date{{\small Institut des Sciences Nucl\'eaires, CNRS-IN2P3/UJF\\
 53, av des Martyrs, 38029 Grenoble Cedex, FRANCE\\
e-mail: Aurelien.Barrau@cern.ch}}

\begin{document}

\baselineskip=.9cm
\bigskip

\maketitle

\bigskip

\begin{center}
{\bf Abstract}\\
\end{center}
The origin of observed extremely high energy cosmic rays remains an astrophysical
enigma. We show that a single evaporating primordial black hole should produce
$8.5 \cdot 10^{14}$ particles over a $10^{20}$ eV threshold. This emission results
from direct production of fundamental constituants and from hadronization of quarks and
gluons. The induced flux on the Earth is studied as a function of the local density
of exploding black holes and compared with experimental data. The discovery potential of
future detectors is finally reviewed.\\

\begin{flushleft}
\end{flushleft}

\bigskip

\begin{center}
 Accepted by {\it Astroparticle Physics}
\end{center}
\newpage

\section{Introduction}

Small Primordial Black Holes (PBHs), with masses well below the self-gravitational collapse limit 
and possibly as low as the Planck Mass ($M_{Pl}\approx5.5\times 10^{-5}g$) may have formed in
the primordial universe \cite{carr}. Numerous processes, compatible with standard
cosmological scenarios, can be put forward to explain their formation \cite{Bousso}. In
particular, if they
result from initial density perturbations (with an initial mass determined by the horizon mass at
this epoch), the mass spectrum can be analytically determined \cite{Gibbon1} following the
natural hypothesis of scale-invariant Gaussian fluctuations:
\begin{equation}\frac{{\rm d}^2n}{{\rm d}M{\rm d}V}=(\alpha
-2)M^{-\alpha}M_{evap}^{\alpha-2}\Omega_{PBH}\rho_{crit}\end{equation} where
$M_{evap}\approx 10^{15}$ g is the mass of a PBH evaporating nowadays, $\Omega_{PBH}$ is the
current density of PBHs in units of critical density $\rho_{crit}$ and
$\alpha=(1+3\gamma)/(1+\gamma)+1$, $\gamma=p/\rho$ being the pressure to density
ratio.

This study is dedicated to the final-stage emission of PBHs to investigate if they can be 
considered as candidates for extremely high energy cosmic rays (EHECR), beyond
100 EeV ($10^{20}$ eV). Observational data \cite{Medina} show that the cosmic ray flux seems
to be curiously unaffected by the expected Greisen-Zatsepin-Kuz'min (GZK) cutoff (due to
interaction with the 2.7K cosmological background above photoproduction threshold). The 
integrated emission of PBHs is estimated in the following sections for a volume of universe 
where predicted effects of this interaction are weak ({\it i.e.} for a radius close to the 
attenuation length $\approx$ 50 Mpc).

\section{Individual emissions}

Hawking showed \cite{Hawking} that black holes can radiate particles in a process qualitatively
equivalent to $e^+e^-$ pairs production in a strong electric field. When the hole temperature
becomes greater that the quantum chromodymanics confinment scale ($T>\Lambda_{QCD}$), {\it
i.e.}
some hundreds of MeV, emitted particles are fundamental constituents rather than composite
hadrons \cite{Gibbon2}. The EHECR production by PBHs (which are particularly affected by evaporation effects
because of their low mass) has to be understood in such an
approach. The emission spectrum for particles of energy $Q$ per unit of time $t$ is:
\begin{equation}\frac{{\rm d}^2N}{{\rm d}Q{\rm d}t}=\frac{\Gamma_s}{h\left(exp\left(\frac{Q}{h\kappa/4\pi^2c}\right)-(-1)^{2s}\right)}\end{equation}
where contributions of angular velocity and electric potential have been neglected since the
black hole discharges and finishes its rotation much faster than it evaporates \cite{Gibbons} 
\cite{Page1}. $\kappa$ is the surface gravity, $s$ is the spin of the emitted species and 
$\Gamma_s$ is the absorption probability. In the general case, $\Gamma_s$ is a
function of $Q$, the particle mass $m$, the hole mass $M$, and the number of degrees of freedom of the
species. Its value can only be computed by numerical approximations based on expansion in
spherical harmonics of the scattering matrix \cite{Press}. In the optical limit ({\it i.e.}
$Q\rightarrow \infty$)
, which is
totally justified for energies considered in this work, 
\begin{equation}\Gamma_s \approx \frac{27Q^2}{64\pi^2 (kT)^2}\end{equation} where $T$ is the "temperature" defined by 
\begin{equation}kT=\frac{hc^3}{16\pi^2GM}\approx 10^{4}\left(\frac{1{\rm g}}{M}\right)~{\rm EeV}\end{equation} In such a description,
the black hole behaviour mimics a black body whose temperature increases when the mass decreases
until it reaches the Planck limit where this theoretical description becomes unadapted. The time-evolution of the system depends on the emitted constituents' degrees of freedom and
is therefore based on the choice of a particle physics model. It is likely
that new particles, absent from the standard model, are emitted when the black hole
temperature becomes extremely high, but the general behaviour remains unchanged: all the
emission above 100 EeV is nearly instantaneous. The mass loss rate of a PBH is 
\cite{Page2} \cite{Halzen}:
\begin{equation}\frac{{\rm d}M}{{\rm d}t}\approx -\frac{(7.8d_{s=1/2}+3.1d_{s=1})\cdot 10^{24}~{\rm
g^3 s}^{-1}}{M^2}\end{equation}
where $d$ is the mass-dependant number of degrees of freedom for the emitted particles of spin $s$.
In the standard model, ${\rm d}M/{\rm d}t\approx -7.9\times10^{26}/M^2$ above the top quark production
threshold. It leads to
\begin{equation}{\rm d}t=\frac{1}{(7.8d_{s=1/2}+3.1d_{s=1})}\cdot\frac{h^3c^6}{(4\pi)^6G^3}\cdot\frac{d(kT)}{(kT)^4}\end{equation}
or
\begin{equation} {\rm d}t_*\approx 1.5 \cdot 10^{-15} \frac{{\rm d}(kT_*)}{(kT_*)^4} \end{equation} where $t_*=t/1{\rm s}$  and $kT_*=kT/1{\rm EeV}$.
Since it has been checked that only particles emitted at $kT\geq 5$ EeV will contribute 
(within a few percent) 
to the flux of cosmic rays with energies beyond 100 EeV, the characteristic production time is $\Delta
t \leq 4\times10^{-18}$ s. As a comparison, the total evaporation time for a $10^{15}$~g
black hole is of the order of the age of the universe.

\section{Extremely high energy emission}

\begin{center}
\begin{figure}[hhh]
$$
\epsfxsize=13cm
\epsfbox{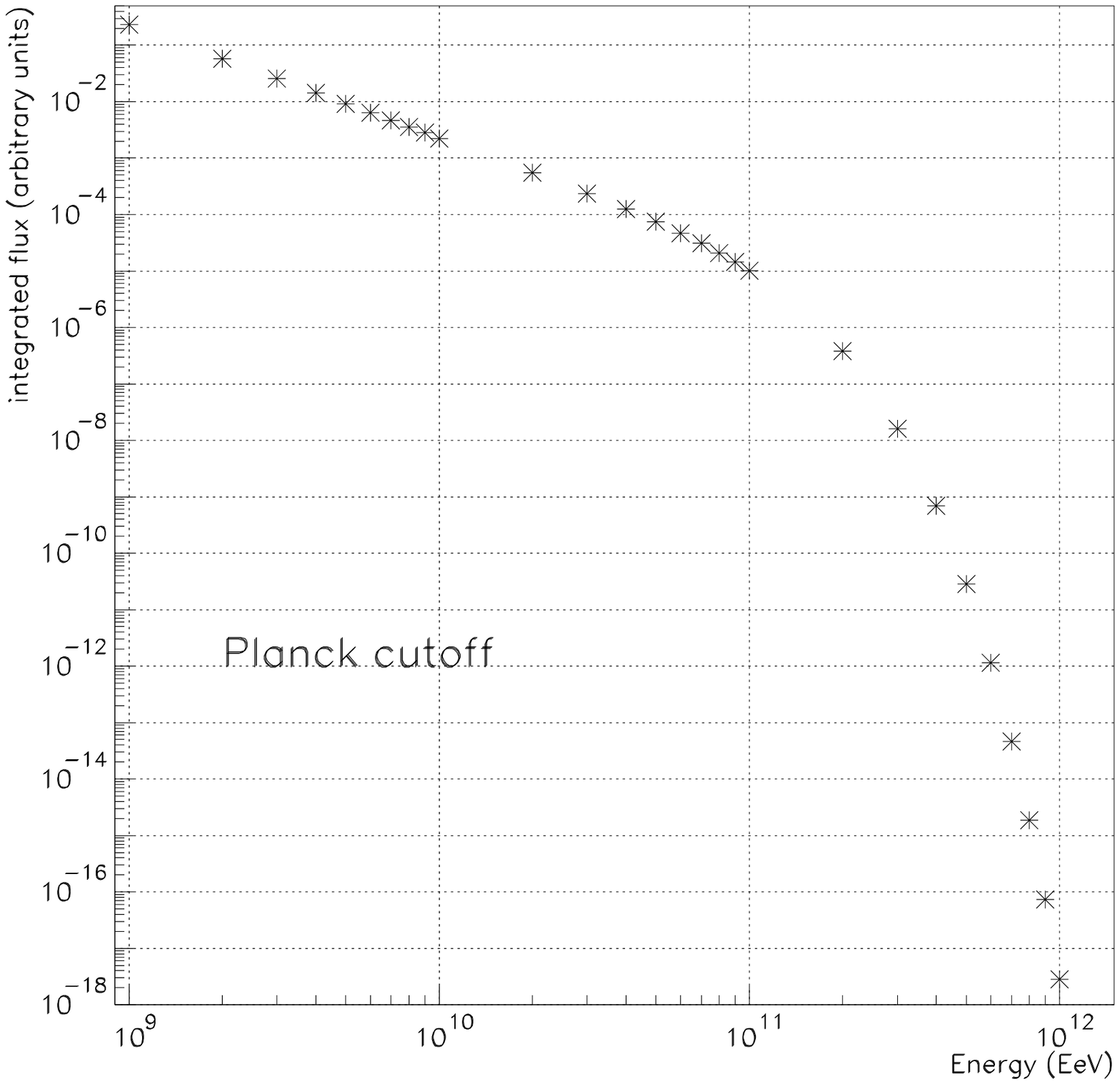}
$$
\caption{Planck cutoff effect on the integrated primary spectrum}
\label{fig:planck}

\end{figure}
\end{center}

Taking into account formula (6) relating the temperature to the mass,
the previous emission spectrum can be rewritten in its integral form \cite{Semikov} per particle 
species above a threshold $E_{th}$:
\begin{equation}N=\frac{1}{(7.8d_{s=1/2}+3.1d_{s=1})}\cdot\frac{27h^2c^9}{8^6\pi^8G^3}\cdot\int_{kT_i}^{kT_{Pl}}
\frac{1}{(kT)^6}\int_{E_{th}}^{\infty}\frac{Q^2 {\rm d}(kT) {\rm d}Q}{e^{Q/(kT)}-(-1)^{2s}}\end{equation}
where $T_i$ and $T_{Pl}$ are the initial and Planck temperatures.
It can be numerically expressed as:
\begin{equation}N=1.56\cdot 10^{16} \int_{kT_{i*}}^{kT_{Pl*}} \frac{{\rm d}(kT_*)}{(kT_*)^3}
\int_{E_{th}/(kT)}^{\infty}\frac{x^2 {\rm d}x}{e^x-(-1)^{2s}}\end{equation}
where $x=Q/(kT)$. Fig \ref{fig:planck} shows that the Planck cutoff is effective
for energies well beyond those of interest here.

\begin{center}
\begin{figure}[hhh]
$$
\epsfxsize=13cm
\epsfbox{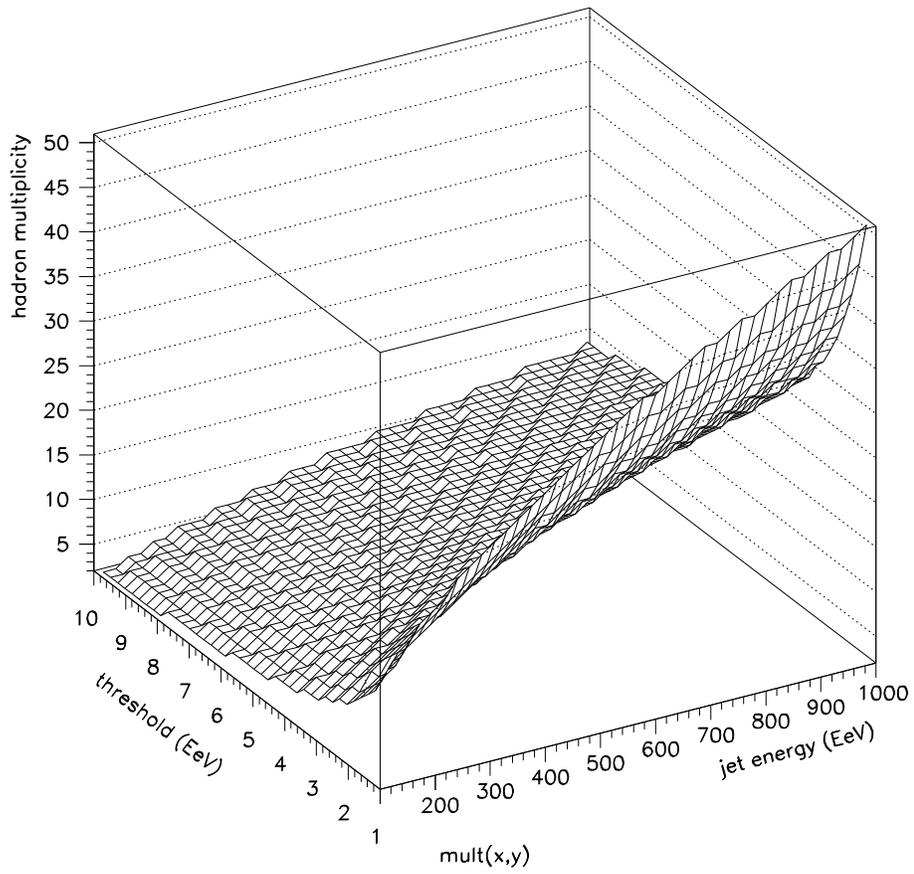}
$$
\caption{Hadron multiplicity as a function of the detection threshold and of the jet energy}
\label{fig:mult}

\end{figure}
\end{center}

After their production, emitted quark and gluons fragment and produce a subsequent number of hadrons. Monte Carlo 
simulation codes tuned to reproduce experimental data obtained on colliders cannot be
used because the energies considered here are several orders of magnitude greater than those
available today. The multiplicity $n_h$ of charged hadrons produced in a jet of energy $Q$ is therefore 
estimated by means of the leading log QCD computation \cite{Bassetto}:
\begin{equation}n_h(Q)\approx 3\times 10^{-2} e^{2.7\sqrt{ln(Q/\Lambda)}}+2\end{equation}
To get the resulting hadron spectrum, Hill \cite{Hill} derived the following distribution:
\begin{equation}\frac{dn_h}{dz}\approx 10^{-1}
e^{2.7\sqrt{ln\frac{1}{z}}}\times\frac{(1-z)^2}{z\sqrt{ln\frac{1}{z}}}\end{equation}
The number of emitted hadrons above the threshold, by a PBH of temperature $T$, can then be written 
as:
\begin{equation}N_h=1.56\cdot 10^{16} \int_{kT_{i*}}^{kT_{Pl*}} \frac{d(kT_*)}{(kT_*)^3}
\int_{mc^2/(kT)}^{\infty}\frac{x^2 dx}{e^x-(-1)^{2s}} \int_{E_{th}/(xkT)}^1 \frac{{\rm
d}n_h}{{\rm d}z}{\rm d}z\end{equation}
per particle species of mass $m$.
The numerical computation has been compared to what is given by the empirical function \cite{Halzen}
$dn_h/dz=(15/16)\times z^{-3/2}(1-z)^2$, leading to a multiplicity which can be
easily calculated analytically. Results are in agreement within an error of 12\% which is certainly
not the dominant uncertainty in this evaluation. Figure \ref{fig:mult} illustrates the general
behavior of the total hadronic multiplicity $\int_{E_{th}/Q}^{1}\frac{dn_h}{dz}dz$ above a given
threshold.

The total number of emitted particles above a detector threshold
$E_{th}=100~{\rm EeV}$ can then be
estimated by summing the direct flux (taking into account all the standard model degrees of
freedom) of fundamental stable particles and the fragmentated flux resulting from the previous
computation for coloured objects. The numerical result is $F(\geq 100~{\rm EeV})\approx 8.5\cdot 10^{14}
$ particles over the lifetime of a PBH.

\section{Resulting flux above 100 EeV}

The derivation of the exact resulting spectrum on the Earth is a complete study by itself,
well beyond the scope of this work. It is straightforward to demonstrate that the integrated 
emission goes as $E^{-2}$, which seems quite difficult to conciliate with the cosmic-ray 
experimental data if energy-dependent confinement effects are ignored. The
following section therefore aims at evaluating the orders of magnitude involved.
\\

To derive the resulting flux reaching the earth, PBHs have been considered as classical (non baryonic)
cold dark matter clustered in galactic halos. The Milky Way mass distribution is therefore assumed
to follow the simple law in spherical coordinates:
\begin{equation}\rho(R)=\rho_\odot \frac{R_c^2+R_\odot^2}{R_c^2+R^2}\end{equation}
where $R$ is the distance between the considered PBHs and the Galactic Center, $\rho_\odot$ is the
local density of exploding PBHs, and $R_c$ is the core radius of the halo.
The particle flux becomes:
\begin{equation}\left(\frac{{\rm d}N}{{\rm d}t}\right)_{galactic}=\rho_\odot \times F\times
J(R_H,R_c,R_\odot)\end{equation}
where \begin{equation}J(R_H,R_c,R_\odot)=\frac{1}{8\pi}
\int_0^\pi \int_0^{R_H} R^2 \frac{R_c^2+R_\odot^2}{R_c^2+R^2} \frac{sin\phi~{\rm
d}R~{\rm d}\phi}
{R_\odot^2-2RR_\odot cos\phi + R^2}\end{equation}
$R_H$ being the total radius of the halo. Table \ref{tab:flux} gives fluxes normalized to the
average for extreme values of $R_c$ and $R_H$ (for $R_\odot=8$ kpc): it shows a quite low 
dependance on the halo parameters.

\begin{table}[htbp]
\begin{center}
\begin{tabular}{|p{2.6cm}|*{4}{c|}|}
\hline
Relative fluxes& $R_H$=40 kpc & $R_H$=100 kpc & $R_H$=200 kpc \\
\hline
$R_c$=2 kpc & 0.97 & 1.01 & 1.03 \\
\hline
$R_c$=4 kpc & 0.94 & 0.99 & 1.01 \\
\hline
$R_c$=6 kpc & 0.96 & 1.02 & 1.05 \\
\hline
\end{tabular}
\end{center}
\caption{Relative fluxes for different halo parameters}
\label{tab:flux}
\end{table}

The extragalactic contribution is computed by assuming a standard galaxy distribution
$\rho_G\approx0.01e^{\pm0.4}h^3~{\rm Mpc^{-1}}$ \cite{Peeb} (with the Hubble parameter defined as 
$H_0 = h\cdot 100$~km.s$^{-1}$.Mpc$^{-1}$). The resulting flux is:
\begin{equation}\left(\frac{{\rm d}N}{{\rm d}t}\right)_{extragalactic}= K(F,R_H,R_c,R_\odot) \times \rho_\odot 
\times \rho_G  \times R_{GZK}\end{equation}
where $K(R_H,R_c,R_\odot)$ is the average emission of a single galaxy (obtained by the previous
method) and $R_{GZK}\approx50~{\rm Mpc}$ is the radius of a sphere "unaffected" by the 
GZK cutoff. On such distances, it is not necessary to redshift energies
within the expected accuracy of a few percent.

Numerical results for average values of physical parameters show that the galactic contribution is
nearly three orders of magnitude larger that the extra-galactic component, even assuming the
highest galaxies number density and the upper Hubble parameter limits (h$\le$1). As it only depends linearly on $R_{GZK}$, the accurate
determination of this radius is also irrelevant.
The total flux above 100 EeV is then: 
\begin{equation}\left(\frac{{\rm d}n}{{\rm d}t}\right)_{PBH}=3.8\cdot 10^{-23} \times \left(\frac{\rho_\odot}{1~{\rm year}^{-1}{\rm
pc}^{-3}} \right)~{\rm m^{-2}s^{-1}sr^{-1}}\end{equation}\\

Experimental data on EHECR show an integrated flux of the order of
$\left(\frac{{\rm d}n}{{\rm d}t}\right)_{exp}\approx 10^{-16}~{\rm m^{-2}s^{-1}sr^{-1}}$ \cite{Takeda}. The
required density of exploding PBHs near the earth to reproduce such a signal
is then $\rho_{\odot}\approx2.6\cdot 10^{6}~{\rm
year}^{-1}{\rm pc}^{-3}$. 

\section{Discussion}

Direct observational constraints on the local PBH explosion rate $\rho_{\odot}$ are
quite difficult to obtain. A reliable search for short bursts of ultra high-energy gamma
radiations from an arbirary direction have been performed using the CYGNUS
air-shower array \cite{Alex}. No strong 1 second burst was observed and the
resulting upper limit, based on the exhaustive analysis of a very fine binning of
the sky, is in the range $\rho_{\odot}\leq 0.9\cdot10^6{\rm year}^{-1}{\rm pc}^{-3}$.
Very similar results were derived by the Tibet \cite{Ame} and the AIROBIC 
collaborations \cite{Funk}. TeV gamma-rays have also been used to search for short
time-scale coincidence events, thanks to the imaging atmospheric Cherenkov technique
developped by the Whipple collaboration. The very high-energy gamma-ray bursts detected
are compatible with the expected background, within $\pm 1.7 \sigma$. The 
resulting upper limit obtained with 5 years of data \cite{Con},  i.e.
$\rho_{\odot}\leq 3\cdot10^6{\rm year}^{-1}{\rm pc}^{-3}$, is substantially 
better than the previous published results in the TeV range. All those limits are
roughly compatible with the density required to generate the
observed EHECR spectrum. 

At the opposite, low-energy ($<0.5$ GeV) cosmic-ray antiprotons
detected by a BESS 13-hours ballon flight have been used \cite{Maki} to put a much more severe 
upper limit of $\rho_{\odot}\leq 2\cdot10^{-2}{\rm year}^{-1}{\rm pc}^{-3}$, which could exclude 
PBHs as serious candidates for EHECR. This analysis is particularly
promising since the authors have shown that the local PHB-antiproton flux can only be due to
contributions from black holes that are very close to explosion, and exist within a few kpc away
from the Solar system. Nevertheless, such data suffer from an important lack of statisics
and from contamination effects due to interactions with the atmosphere. Future results from the AMS
\cite{AMS} spectrometer on board the International Space Station will give a much more accurate
antiproton spectrum in the 0.1-1 GeV range. Those data should allow a stringent upper limit
(if not a positive detection) on nearby exploding PBHs.\\

An entirely different approch is to study the diffuse gamma-ray background spectrum. The emission
from PBHs over the lifetime of the Universe is integrated so as to evaluate the resulting particles
and radiations. This method \cite{Halzen} leads to 
$\rho_{\odot}\leq 10 {\rm year}^{-1}{\rm pc}^{-3}$ for clustered black holes. It should, anyway, be
emphasized that such a study does not directly constrain $\rho_{\odot}$. The
resulting "Page-Hawking bound" \cite{Page3} on $\Omega_{PBH}$,
derived to match the observed spectrum at 100 MeV, is converted into an upper limit on the initial
number density of holes per logarithmic mass interval $N_{PBH}$ at $M=M_{evap}$ under assumptions on the
Hubble parameter, on the relative matter density ($\Omega_M$), on the equation of state of the
Universe at the formation epoch, and on the gaussian distribution of initial density
perturbations. This latter point is rather controvertial. The upper limit on  $\rho_{\odot}$ which
can then be derived has to account for the large (possibly up to 8 orders of magnitude) uncertainties 
associated with clustering. Recent reviews on the detection of PBHs captured around massive objects 
\cite{Deri} show that, when the first astrophysical objects with masses of the order of the
Jeans mass were forming, black holes haloes on the sub-galactic scale could have formed around old
globular clusters, dark matter clusters or population III stars. This makes the use of 100 MeV
gamma-rays a quite difficult way of ruling out an important local rate of PBH
explosions, though future GLAST \cite{GLAST} data should change the situation by a
dramatic improvement in statistics and resolution.\\

Furthermore, the first results from the AUGER
Observatory \cite{Auger} will soon give high statistics samples of EHECR. With the PBH space
distribution previously assumed, the resulting EHECR flux would be from 6.0 to 2.2 times higher
in the Galactic Center direction than in the opposite direction, for an integrated observation angle from
10 to 90 degrees. After five years of operation, the AUGER observatory should collect up to 300
cosmic rays above 100 EeV. Such an anisotropy would be detectable if more than 
approximately 50\% of them come from PBHs.
\\

Finally, some evidences for PBH signatures available nowadays should be noted.
Studies of the BATSE 1B and 1C catalogs have shown \cite{Cline1} that some gamma-ray bursts (GRBs) 
were consistent with a PBH evaporation origin at the quark-gluon phase transition. 
Characteristics of selected events are in remarkable agreement with the "Fireball" PBH picture. The
resulting (model dependent) limit is significantly lower than what is expected in the
present work, and disfavours a PBH
origin for EHECR. Nevertheless, new analysis of EGRET data \cite{Cline2} gives some
evidences for a gamma-ray halo "glow" due to PBH emission. Those first tentative detections are very
promising for further investigations on the subject.
\\

From the theoretical point of view, it should also be emphasized that results given in this paper 
are based on the standard particle physics model. The probable increase of degrees of freedom 
available when the black hole temperature exceeds energies currently available on colliders would
modify the estimated fluxes, making the final explosion much more violent. This could validate
PBHs as a good source candidate for a fraction of the observed high energy cosmic rays.
\\

{\bf Acknowledgments}. I would like to thank Cecile Renault for very helpful discussions.

\newpage

\newpage

\end{document}